\documentclass[runningheads,a4paper]{llncs}

\usepackage[cmex10]{amsmath}
\usepackage{amssymb}
\usepackage{llncsdoc}
\usepackage{mathptmx}       
\usepackage{helvet}         
\usepackage{courier}        
\usepackage{makeidx,cite}         
\usepackage[pdftex]{graphicx}
\usepackage{epstopdf}
\DeclareGraphicsExtensions{.pdf,.jpeg,.png,.eps}
\usepackage{multirow}        
\usepackage[bottom]{footmisc}

\usepackage{url}
\usepackage[linesnumbered,ruled,vlined]{algorithm2e}
\usepackage{array}
\usepackage{booktabs}
\usepackage[caption=false,font=footnotesize]{subfig}
\usepackage{xcolor,colortbl}
\usepackage{bm}
\usepackage{subfig}

\DeclareMathSymbol{\R}{\mathalpha}{AMSb}{"52}

\definecolor{Gray}{gray}{0.85}
\definecolor{LightCyan}{rgb}{0.88,1,1}

\begin{document}
\mainmatter 

\title{A Wavelet Diffusion GAN for Image Super-Resolution}

\author{Lorenzo Aloisi \and Luigi Sigillo \and Aurelio Uncini \and Danilo Comminiello
\thanks{
This work was partly supported by “Ricerca e innovazione nel Lazio - incentivi per i dottorati di
innovazione per le imprese e per la PA - L.R. 13/2008” of Regione Lazio, Project ``Deep Learning Generativo nel Dominio Ipercomplesso per Applicazioni di Intelligenza Artificiale ad Alta Efficienza Energetica”, under grant number 21027NP000000136 and by the ``Progetti di Avvio alla Ricerca", under grant AR123188B4D1FCB5 and by the European Union under the Italian National Recovery and Resilience Plan (NRRP) of NextGenerationEU, ``Rome Technopole" (CUP B83C22002820006)—Flagship Project 5: ``Digital Transition through AESA radar technology, quantum cryptography and quantum communications". The work of A. Uncini has been partly supported by the European Union under the NRRP of NextGenerationEU, partnership on ``Future Artificial Intelligence Research” (PE0000013 - FAIR - Spoke 5: High Quality AI).
}
}
\titlerunning{A Wavelet Diffusion GAN for Image Super-Resolution}
\institute{Department of Information Engineering, Electronics and Telecommunications (DIET), \\ ``Sapienza'' University of Rome, \\ Via Eudossiana 18, 00184, Rome. \\Email: \{luigi.sigillo; aurelio.unicini; danilo.comminiello\}@uniroma1.it}

\maketitle

\begin{abstract}
In recent years, diffusion models have emerged as a superior alternative to generative adversarial networks (GANs) for high-fidelity image generation, with wide applications in text-to-image generation, image-to-image translation, and super-resolution. However, their real-time feasibility is hindered by slow training and inference speeds. This study addresses this challenge by proposing a wavelet-based conditional Diffusion GAN scheme for Single-Image Super-Resolution (SISR). Our approach utilizes the diffusion GAN paradigm to reduce the timesteps required by the reverse diffusion process and the Discrete Wavelet Transform (DWT) to achieve dimensionality reduction, decreasing training and inference times significantly. The results of an experimental validation on the CelebA-HQ dataset confirm the effectiveness of our proposed scheme. Our approach outperforms other state-of-the-art methodologies successfully ensuring high-fidelity output while overcoming inherent drawbacks associated with diffusion models in time-sensitive applications. The code is available at \url{https://www.github.com/aloilor/WaDiGAN-SR}
\end{abstract}
\section{Introduction}
\label{sec:intro}
Diffusion models have surfaced as a robust solution for high-fidelity image generation and proved that they can beat the previous state-of-the-art paradigm headed by generative adversarial networks (GANs)\cite{DDPM, ddpmvsgan}. They provide more training stability and a high degree of flexibility in handling conditional generation. This versatility in handling diverse conditional inputs allows for their integration into a broad range of applications, such as text-to-image generation \cite{tumanyan2023plug, zhang2023adding}, image-to-image translation \cite{zeroshot, sigillo2024ship, 10181838}, image inpainting \cite{wang2023imagen}, and image restoration \cite{li2023efficient} among others. 

Moreover, diffusion models have also been introduced into the domain of Image Super-Resolution with a conditional diffusion-based approach \cite{SR3}. However, their slow training, due to the diffusion process being held in the pixel space, and inference speeds pose a significant limitation on their application in real-time scenarios. 
Diffusion models based on pixel diffusion \cite{DDPM,SR3} require thousands of steps to produce a high-quality sample and take several minutes to generate a single image. 
On the other hand, another family of approaches in the literature is the latent space diffusion one \cite{9878449} which tries to smooth out the drawbacks of classical diffusion models, but such models rely on large-scale datasets to train a variational autoencoder (VAE), to provide a latent space on which to perform the diffusion process on. 
To take advantage of the limitations associated with the training of latent diffusion models in image super-resolution in \cite{stablesr} they leverage Stable Diffusion \cite{9878449} to explore the prior given by the pretrained encoder, aiming to reduce inference time while maintaining high-quality output. However, they still exhibit a relatively high runtime.
Diffusion GAN \cite{DDGAN} is an alternative to avoid doing such intense large-scale training, achieving a breakthrough in accelerating the inference speed of unconditional image generation by integrating diffusion and GANs within a unified system. This innovative approach results in a substantial reduction of inference to a few seconds. 

WaveDiff \cite{WaveDiff} achieved an even further breakthrough by introducing a wavelet approach into unconditional image generation, using the idea of Diffusion GAN \cite{DDGAN}, resulting in an additional speed-up in both training and inference speeds. 
Our study seeks to investigate the challenges of a faster diffusion on super-resolution problems, by proposing a novel wavelet-based conditional Diffusion GAN approach.
Our solution relies on the Discrete Wavelet Transform (DWT) applied to both the image and feature levels, a method that breaks down each input into four sub-bands representing low and high-frequency components. The wavelet transform allows for spatial dimension reduction, according to the Nyquist Rule \cite{nyquist}, and increased image detail due to the leveraging of high-frequency information.
Incorporating a hybrid approach between diffusion and GAN and by taking advantage of wavelet sub-bands, we can achieve a significant reduction in both training and inference times while outperforming state-of-the-art quality and maintaining high-fidelity output. To summarize, our work makes the following contributions: 
\begin{itemize}
    \item We propose the first Wavelet-based conditional Diffusion GAN approach to Image Super-Resolution that exploits the spatial dimension reduction and the enhanced image detail provided by leveraging high-frequency information through the wavelet sub-bands.
    \item Our method outperforms previous state-of-the-art for face super-resolution in terms of training and inference times and image quality, serving as a stepping stone for real-time applications of diffusion models. 
\end{itemize}

The work is divided into different sections, in section \ref{sec:background} we introduce some relevant related works, in \ref{sec:proposed_method} we show and explain the method proposed, while in \ref{sec:exp_results} we exhibit the experimental results obtained and finally in \ref{sec:conclusion} we draw the conclusions.

\section{Background}
\label{sec:background}
Wavelet transform is a mathematical tool used for signal processing and image analysis. It decomposes a signal into components with different frequency bands. Wavelet analysis allows for a localized and multiresolution representation of images, making it particularly well-suited for capturing both fine details and global features. The images are passed through a high pass and a low pass filter and then decomposed into high-frequency and low-frequency components.
Several approaches experiment in the domain of wavelet transforms for computer vision tasks \cite{guth2022wavelet, xue2024low} and also medical scenarios \cite{10268972,sigillo2024generalizing}. In \cite{SWAGAN} they propose a style and wavelet-based GAN, integrating wavelets into both the generator and discriminator architectures, ensuring that the latent representations at each stage of the generation process are sensitive to frequency components. Another step in \cite{guth2022wavelet} is the introduction of a wavelet score-based generative model to mitigate the high computational costs required to solve the SDEs used to generate data samples. WaveDM \cite{WaveDM} instead is an approach that mixes wavelet and diffusion models to perform image restoration, which also introduced an efficient sampling strategy to reduce inference and training times.
Only in \cite{DiWa} they show a first approach to the problem of super-resolution using diffusion-Wavelet (DiWa), hallucinating high-frequency information for super-resolved images on the wavelet spectrum.
Although excluding \cite{DiWa}, none of those above methods address the task of image super-resolution, they demonstrate the effectiveness of incorporating high-frequency information to improve image detail and reduce computational complexity through wavelet sub-band utilization. Indeed this problem is often solved in literature with the aid of GANs \cite{esrgan, wang2021real}. In contrast, our methodology aims to overcome the inherent challenges of image super-resolution, specifically the protracted inference and training periods resulting from the utilization of a diffusion model. This is achieved through the incorporation of the wavelet transform within a conditional Diffusion GAN architecture.

\section{Proposed method}
\label{sec:proposed_method}

\begin{figure*}[h]
    \centering
    \includegraphics[width=\linewidth]{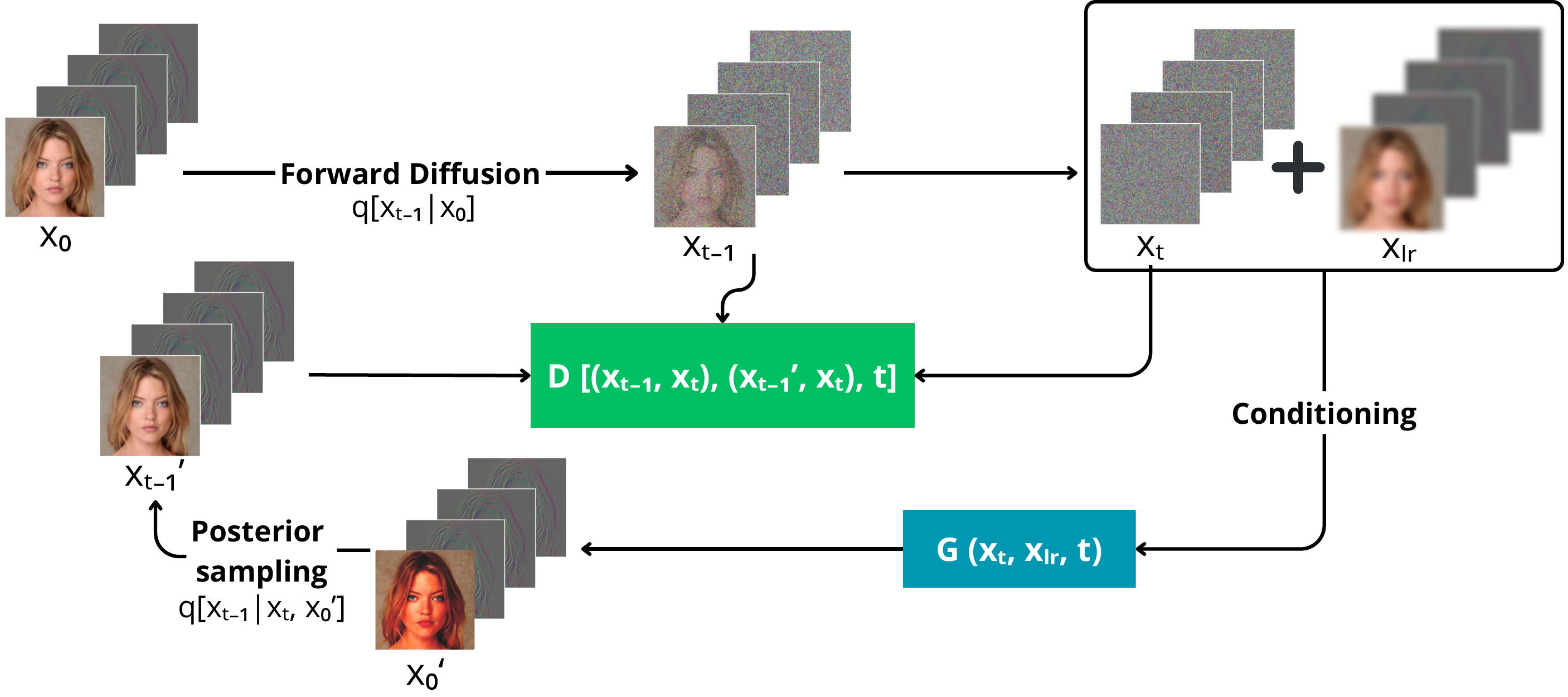}
    \caption{Method architecture and training scheme. In green our discriminator and in blue our conditional generator. \(x_0\) undergoes forward diffusion in wavelet space and the resulting pure noise \(x_t\) gets concatenated to the low-res input \(x_{lr}\) to condition the generator for the backward diffusion.}
    \label{fig:cond_diff_scheme}
\end{figure*}

We allow larger step sizes in the reverse diffusion process by using a GAN \cite{DDGAN}. 
We use the low-resolution images as conditioning input to the model, but we perform the whole diffusion process in wavelet space instead of pixel space. 

\subsection{Discrete Wavelet Transform}

Wavelet transform decomposes signals into localized frequency bands, providing both time and frequency localization. Unlike the Fourier transform, Wavelet transform can capture local changes in frequency over time, enabling a more flexible representation. Through multiresolution analysis, wavelets capture both fine details and global trends, making them effective in tasks like denoising and compression. In our approach, the Discrete Wavelet Transform (DWT) highlights high-frequency features and reduces spatial dimensions, thereby improving sampling efficiency. \\
\noindent
Let \( L = \sqrt{\frac{1}{2}} \begin{bmatrix} 1 & 1 \end{bmatrix} \) and \( H = \sqrt{\frac{1}{2}} \begin{bmatrix} -1 & 1 \end{bmatrix} \) denote low-pass and high-pass filters, respectively. These filters are utilized to decompose the high-resolution input image \(x \in \mathbb{R}^{H \times W}\) into four wavelet sub-bands \(X_{ll}\), \(X_{lh}\), \(X_{hl}\), and \(X_{hh}\) with a size of \(\frac{H}{2} \times \frac{W}{2}\). For an input image \(x\) belonging to \(\mathbb{R}^{3 \times H \times W}\) we decompose it into low and high sub-bands, which are subsequently concatenated channel-wise to form a unified target \(x_0\) in \(\mathbb{R}^{12 \times \frac{H}{2} \times \frac{W}{2} }\).

\subsection{Wavelet-based conditional diffusion scheme}
Figure \ref{fig:cond_diff_scheme} illustrates our method. The forward diffusion process follows the traditional definition \cite{DDPM}, but it is done in wavelet space instead of pixel space: noise gets progressively added to \(x_0\), the wavelet decomposition of the high-resolution input, until only pure Gaussian noise \(x_t\) is remaining. This pure Gaussian noise will be concatenated to \(x_{lr}\), the wavelet decomposition of the low-resolution input image, to model the backward diffusion process. 

During the reverse process, our generator, given \(x_t\) and \(x_{lr}\) will produce an unperturbed prediction \(x_0'\) of the initial high-res image, which will be used to take from the tractable posterior distribution the perturbed sample \(x_{t-1}'\). In inference time, this sampling process will be iterated for a predefined number of timesteps T \((\leq 8)\) to complete the denoising process and acquire the super-resolved image. 

The role of the discriminator during the training phase is to distinguish between real pairs \((x_{t-1}, x_t, t)\) and fake pairs \((x_{t-1}', x_t, t)\).

\subsection{Conditional backward diffusion}

\begin{figure*}[h] 
    \centering
    \includegraphics[width=\linewidth]{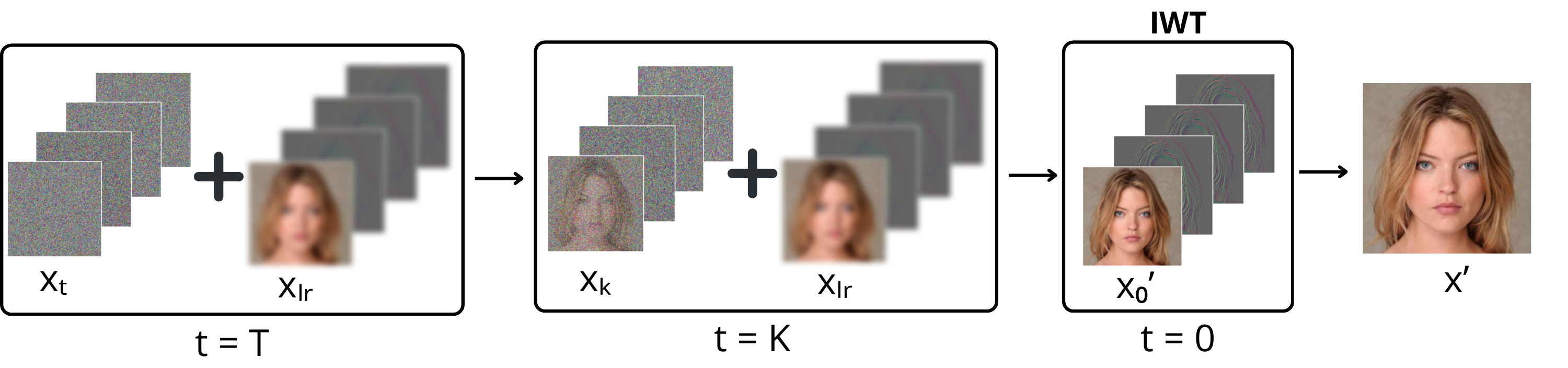}
    \caption{Reverse diffusion process and inference: the model \(p_{\theta}\) iteratively produces a more refined sample from \(x_t\) and \(x_{lr}\). After \(T\) iterations, \(x_0'\) is used to reconstruct the super-resolved image.}
    \label{fig:reverse_diff}
\end{figure*}

Instead of the traditional reverse diffusion approach which requires several thousands of steps to properly denoise the image, the GAN utilization in our architecture enables us to do bigger steps in the diffusion process (Figure 2). This assumption implies that a Gaussian distribution cannot approximate the reverse process but instead, we use a multimodal distribution, which will be implicitly modeled by the time-dependent GAN generator \(G(x_t, x_{lr}, t)\). In this formulation, the model does not directly predict \(x_{t-1}\). Instead, it predicts the clean image \(x_{0}\) and uses the known diffusion process to obtain \(x_{t-1}\). Specifically \(x_{t}\) is the noisy image at timestep \(t\); \(x_{lr}\) is obtained by first scaling the low-resolution input to the target resolution via bicubic interpolation, and then applying a discrete wavelet transform to decompose it into multiple frequency sub-bands; \(f_{\theta}(x_{t}, x_{lr}, t)\) is the neural network (the “denoising generator”) that estimates the clean image \(x_{0}\).\\
Once \(x_{0}\) is estimated, the probability distribution 
\(p_{\theta}(x_{t-1} \mid x_{t}, x_{lr})\) is defined using the standard posterior from the forward diffusion process, but with \(x_{0}\) replaced by the network’s estimate \(f_{\theta}(x_{t}, x_{lr}, t)\). Formally:

\begin{equation}
\label{eq:diffusion_posterior}
p_{\theta}(x_{t-1} \mid x_{t}, x_{lr}) 
:= q\bigl(x_{t-1} \mid x_{t},\, x_{0} = f_{\theta}(x_{t}, x_{lr}, t)\bigr).
\end{equation}

\noindent In other words, the model learns to reconstruct the clean image \(x_{0}\) at each timestep, and the actual sampling step to get \(x_{t-1}\) follows the known diffusion posterior with this estimated \(x_{0}\).

\subsection{Optimization}
Given the time-dependent discriminator $D(x_{t-1}, x_t, t) : \mathbb{R}^N \times \mathbb{R}^N \times \mathbb{R} \rightarrow [0, 1],$   
and the generator $ G(x_t, x_{lr}, t),$ the training process is set up in an adversarial manner to match the conditional GAN generator \(p_{\theta}(x_{t-1}|x_t, x_{lr})\) and the posterior distribution \(q(x_{t-1}|x_t, x_0)\), optimizing both the generator \(G\) and the discriminator \(D\) using adversarial losses: 
 \begin{equation}
       \label{eq:advd}
        L_{adv}^D = -\log(D(x_{t-1}, x_t, t)) + \log(D(x_{t-1}', x_t, t)) 
 \end{equation}
\begin{equation}
      \label{eq:advg}
    L_{adv}^G = -\log(D(x_{t-1}', x_t, t))
\end{equation}
Furthermore, following WaveDiff \cite{WaveDiff}, we use a reconstruction term to fully exploit the wavelet sub-bands. This term serves a dual purpose: preventing the loss of frequency information and maintaining the consistency of wavelet sub-bands. The formulation involves an L1 loss between the generated image and its ground-truth counterpart:
\begin{equation}
      \label{eq:loss_rec}
    L_{rec} = \|x_0' - x_0\|
\end{equation}

\noindent The total objective of the generator will be a linear combination of the adversarial and the reconstruction losses:

\begin{equation}
      \label{eq:loss_tot}
    L^G = L_{adv}^G +  \lambda L_{rec}
\end{equation}

\section{Experimental results}
\label{sec:exp_results}
\textbf{Resources constraints} It is important to note that the neural networks in this study were trained with limited resources, thus some of the considerations in this chapter result from the constraints on training and testing time and data and would benefit from further research and experimentation.
We evaluate the model for face super-resolution to demonstrate our approach's effectiveness, comparing it to several other state-of-the-art approaches. Finally, we perform a small ablation study on a real-world dataset of ships \cite{sigillo2024ship}.

\subsection{Training details}
We train all the models for 25k iteration steps.
\textbf{Dataset} Following \cite{SR3}, we experiment on CelebA-HQ \cite{CelebA}, a large-scale face attributes collection of 30k celebrity images. We employ bicubic interpolation to produce LR-HR image pairs, resizing all images to match the task of 8x magnification factor 16x16 \(\rightarrow\) 128x128.  

\noindent \textbf{Generator architecture} We employed a U-Net-like architecture \cite{unet} for the generator: our implementation is based on WaveDiff \cite{WaveDiff}, but we remove the conditioning on the latent variable \(z\) since it is not needed for our purposes. The wavelet information is integrated into the feature space of the generator: instead of using the standard downsampling and upsampling blocks, frequency-aware blocks and frequency-aware residual connections are employed to enhance the generator's awareness of high-frequency components, resulting in sharper and more detailed images.

\noindent\textbf{Experimental setup} We trained our model on the CelebA-HQ (16, 128) dataset with learning rates set to \(2\times10^{-4}\) and \(1\times10^{-4}\) for the generator and discriminator, respectively. The generator architecture uses base channels of 64, with a channel multiplier following the pattern [1, 2, 2, 2, 4] and employs two ResNet \cite{ResNet} blocks per level. The discriminator consists of 6 layers, and we utilize 2 timesteps in the reverse diffusion process for super-resolution. For optimization, we employ the Adam optimizer \cite{adam} with decay rates \(\beta_1 = 0.5\) and \(\beta_2 = 0.9\), along with an Exponential Moving Average (EMA) with a decay factor of 0.9999. Training was conducted over 25k iteration steps using a batch size of 64 and lazy regulation with a value of 10 to ensure stable training dynamics.


\subsection{Quantitative results}
\textbf{Evaluation metrics} We compare our model to the other baselines using standard metrics for the reconstructed image quality, such as Peak Signal-to-Noise ratio (PSNR), Structural Similarity Index (SSIM), Learned Perceptual Image Patch Similarity (LPIPS) and Fréchet Inception Distance (FID). 
\begin{figure*}[t] 
    \centering
    \includegraphics[width=\linewidth]{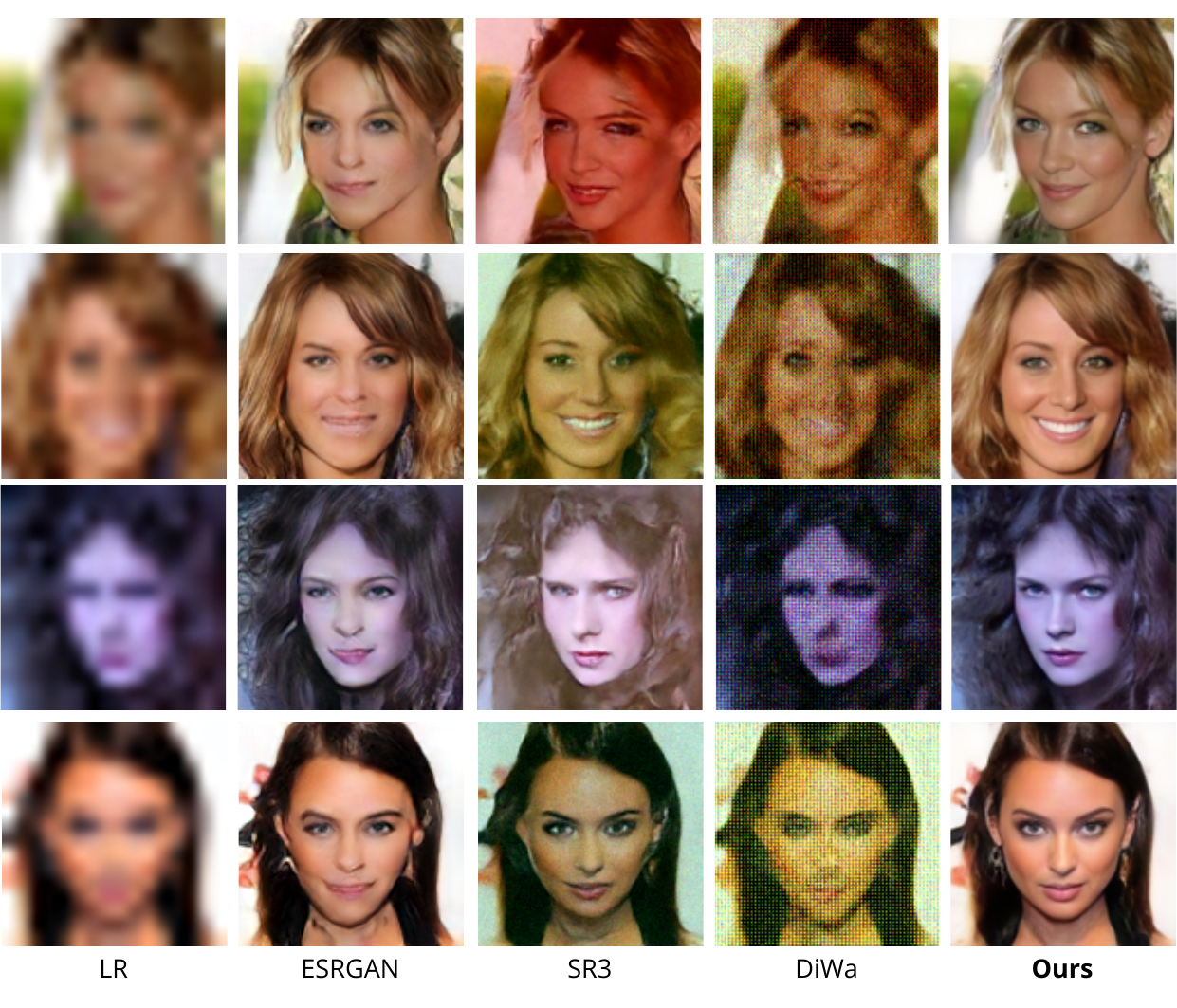}
    \caption{Qualitative comparison between our model, ESRGAN, SR3 and DiWa trained for 25k iteration steps on CelebA-HQ for the task of 16x16 \(\rightarrow\) 128x128.}
    \label{fig:qual_comp}
\end{figure*}
\begin{table}[h]
\centering
\begin{tabular}{lcccc} 
    \toprule
    Metric & ESRGAN \cite{esrgan} 
    & SR3 \cite{SR3}& DiWa \cite{DiWa} & \textbf{Ours} \\
    \midrule
    PSNR \(\uparrow\) & \underline{21.13} 
    & 14.65  & 13.68 & \textbf{23.38} \\
    SSIM \(\uparrow\) & \underline{0.59} 
    & 0.42  & 0.13 & \textbf{0.68}\\
    LPIPS \(\downarrow\) & \underline{0.082} 
    & 0.365 & 0.336 & \textbf{0.061} \\
    FID \(\downarrow\) & \textbf{20.8} 
    & 99.4 & 270 & \underline{47.2} \\
    \midrule
    Runtime & 0.04s & 60.3s & 34.7s & \underline{0.12s} \\
    Parameters & 31M & 98M & 92M & \underline{57M} \\
    \bottomrule
\end{tabular} 
\vspace*{0.2cm} 
\caption{Evaluation metrics comparison for our model, ESRGAN, SR3 and DiWa trained for 25k iteration steps on CelebA-HQ for the task of 16x16 \(\rightarrow\) 128x128.}

\label{tab:quant_metrics}
\end{table}

\noindent
\textbf{Inference time and parameters} We also opted to compare the models in terms of inference time and parameters to show how powerful and lightweight our method is compared to the other baselines. 


\noindent
Table \ref{tab:quant_metrics} demonstrates that, even with a limited number of iterations $(25k)$, our method achieves excellent performances and surpasses the other baselines in the 16x16 \(\rightarrow\) 128x128 task. 
Notably, even though our model has a significantly lower number of parameters compared to the diffusion baselines, we still achieve substantial performance improvements, thus delivering state-of-the-art results with considerably reduced computational complexity and inference time, showing that our approach is suitable for real-time image super-resolution applications. 
\begin{figure*}[!t] 
    \centering
    \includegraphics[width=0.8\linewidth]{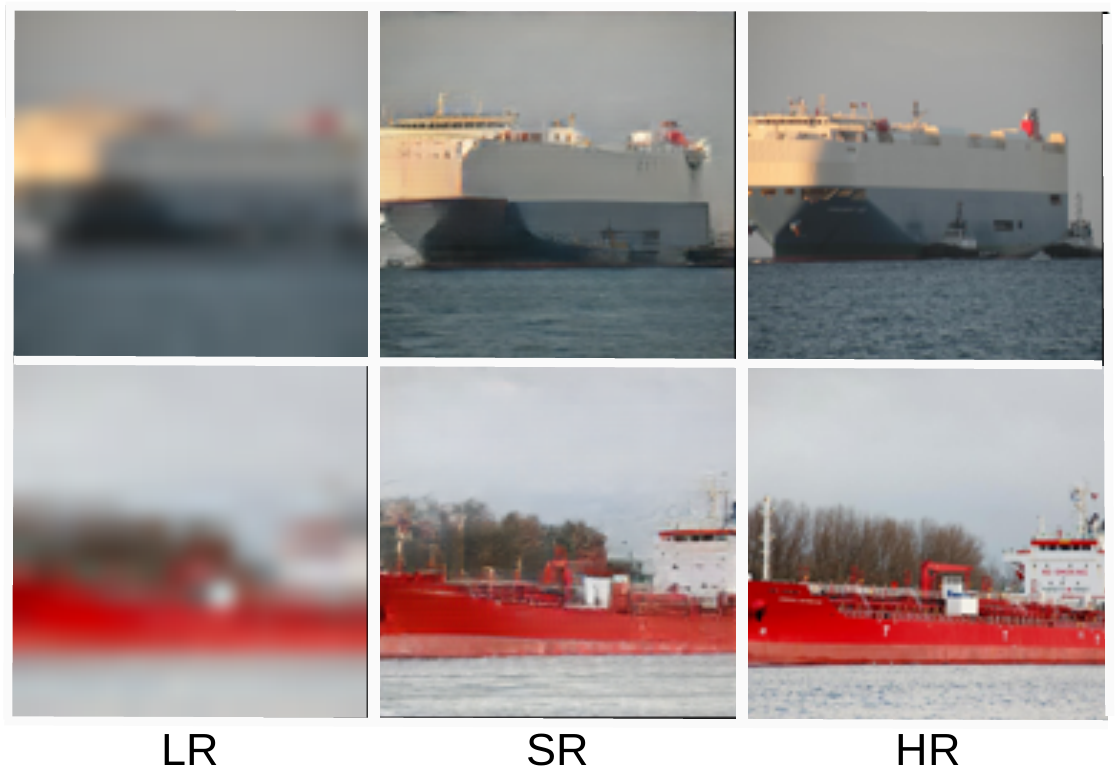}
    \caption{Results on Shipspotting \cite{sigillo2024ship}.}
    \label{fig:qual_comp}
\end{figure*}

\subsection{Qualitative results}
Due to the significant magnification factor (8x), there are numerous possible outputs. Therefore, an exact match with the reference image is not expected.
Figure \ref{fig:qual_comp} reveals that our model excels in enhancing image details, providing a superior visual fidelity compared to the ESRGAN \cite{esrgan}, SR3 \cite{SR3} and DiWa \cite{DiWa} counterparts. The generated images showcase a significant reduction in artifacts, preserving finer details and avoiding common distortions often observed in all of the other baselines. More specifically, the traditional diffusion baselines suffer from annoying crosshatch artifacts, along with color shift artifacts, where the color distribution of the reconstructed image does not correspond with that of the target image, as noted in \cite{color-shift}. 
Notably, even with a number as low as 25k iteration steps, our method provides state-of-the-art results and a high degree of visual fidelity, further proving its suitability for real-time Image Super-Resolution applications and tasks demanding high-fidelity output. 

\noindent We conducted an extensive ablation study utilizing a real-world dataset of ships, specifically the Shipspotting dataset \cite{sigillo2024ship}, to evaluate the performance of our super-resolution framework under practical conditions. This dataset, comprising diverse ship images captured in real-world scenarios, provided a robust testbed for assessing the ability of our model to recover fine-grained details, preserve structural integrity, and generalize across varying environmental conditions.  This rigorous evaluation underscores the practical applicability of our approach in real-world settings, particularly in domains such as maritime surveillance and remote sensing, where high-resolution imagery is critical.

\section{Conclusion}
\label{sec:conclusion}

This study presents a wavelet-based diffusion approach to enhance the practical applicability of diffusion Models in Image Super-Resolution.
The core contributions of this work lie in leveraging both the discrete wavelet transform (DWT) to exploit dimensionality reduction and increased attention to high-frequency details, and a conditional Diffusion GAN architecture to effectively reduce both convergence and inference times while still providing state-of-the-art quality of generated images.
Due to hardware limitations, our experimentation is constrained, and our method would greatly benefit from further research and experimentation on other datasets and a bigger image size to explore its efficacy and performance in diverse settings. 
Nonetheless the experimental results \ref{sec:exp_results} validate the efficiency of our model, demonstrating its superiority over the baselines in terms of many metrics.
It shows also that our method is fast and lightweight (57M parameters) with $0.12$ seconds of inference time.
As a result, our approach is proven to be highly effective and resource-efficient, and it yields state-of-the-art outcomes, providing a solution to the time challenges faced by classical diffusion models, and making them appealing for other researchers operating under hardware constraints.

\bibliographystyle{splncs03}
\bibliography{refs}

\end{document}